\documentclass{article}

    \PassOptionsToPackage{numbers, compress}{natbib}

\usepackage[preprint]{neurips_2023}

\usepackage[utf8]{inputenc} 
\usepackage[T1]{fontenc}    
\usepackage{hyperref}       
\usepackage{url}            
\usepackage{booktabs}       
\usepackage{amsfonts}       
\usepackage{amsmath}
\usepackage{amssymb}
\usepackage{mathtools}
\usepackage{amsthm}
\usepackage{nicefrac}       
\usepackage{microtype}      
\usepackage{xcolor}         

\usepackage{bm}

\usepackage{hyperref}
\usepackage{multirow}
\usepackage[noabbrev,capitalize,nameinlink]{cleveref}
\usepackage{lipsum}  

\usepackage{color}
\usepackage[group-separator={,},group-minimum-digits={3}]{siunitx}
\usepackage{enumitem}
\usepackage[pdftex]{graphicx}
\usepackage{wrapfig}

\usepackage{musicography}

\title{MuseCoco: Generating Symbolic Music from Text
}

\author{
  Peiling Lu$^{\ddagger*}$, Xin Xu$^{\fl*}$, Chenfei Kang$^{\sh*}$, Botao Yu$^{\na*}$, Chengyi Xing$^{\musDoubleSharp}$\thanks{These authors contributed equally to this work.}\ \ , Xu Tan$^{\ddagger}$\thanks{Correspondence: Xu Tan, xuta@microsoft.com}\ \ ,  Jiang Bian$^{\ddagger}$ \\
  $^{\ddagger}$Microsoft Research Asia \\
  $^{\fl}$Zhejiang University, $^{\sh}$Shanghai Jiao Tong University \\
  $^{\na}$Nanjing University, $^{\musDoubleSharp}$Dalian University of Technology \\
  \texttt{\{peil, xuta, jiabia\}@microsoft.com} \\
  \texttt{xxucs@zju.edu.cn, chenfeikang314@gmail.com} \\
  \texttt{btyu@foxmail.com, xcyhbp@mail.dlut.edu.cn} \\
  \url{https://github.com/microsoft/muzic}
}

\begin{document}
\maketitle

\begin{abstract}
Generating music from text descriptions is a user-friendly mode since the text is a relatively easy interface for user engagement. 
While some approaches utilize texts to control music audio generation, editing musical elements in generated audio is challenging for users.
In contrast, symbolic music offers ease of editing, making it more accessible for users to manipulate specific musical elements. In this paper, we propose MuseCoco, which generates symbolic music from text descriptions with musical attributes as the bridge to break down the task into text-to-attribute understanding and attribute-to-music generation stages. 
MuseCoCo stands for Music \textbf{Co}mposition \textbf{Co}pilot that empowers musicians to generate music directly from given text descriptions, offering a significant improvement in efficiency compared to creating music entirely from scratch.
The system has two main advantages: Firstly, it is data efficient. In the attribute-to-music generation stage, the attributes can be directly extracted from music sequences, making the model training self-supervised. In the text-to-attribute understanding stage, the text is synthesized and refined by ChatGPT based on the defined attribute templates. Secondly, the system can achieve precise control with specific attributes in text descriptions and offers multiple control options through attribute-conditioned or text-conditioned approaches. 
MuseCoco outperforms baseline systems in terms of musicality, controllability, and overall score by at least 1.27, 1.08, and 1.32 respectively. Besides, there is a notable enhancement of about 20\% in objective control accuracy. In addition, we have developed a robust large-scale model with 1.2 billion parameters, showcasing exceptional controllability and musicality. Music samples generated by MuseCoco are available via this link \footnote[1]{\url{https://ai-muzic.github.io/musecoco/}}, and the code is available at this link \footnote[2]{\url{https://github.com/microsoft/muzic/musecoco/}}.
\end{abstract}

\section{Introduction}

Text-to-music generation is an important task in automatic music generation because it allows users to generate music more easily and intuitively, using natural language as an interface. This makes it a user-friendly mode of music generation, particularly for those who do not have a background in music theory or composition. Previous work \cite{agostinelli2023musiclm, huang2023noise2music, Erniemusic, museai} generating musical audio from texts faces the following challenges: 1) Limited adaptability: musical audio generated from texts is less adaptable than symbolic music. Once the audio is produced, it may be difficult to make significant changes to the music without starting the generation process over. 2) Lack of control: generating musical audio from text descriptions lacks control over specific aspects of the music such as tempo, meter, and rhythm since the generation process is not explicitly controlled. However, they can be easily handled by generating symbolic music from given texts. Symbolic music refers to a type of music notation that uses symbols and music language to represent specific music ideas, thus, it is more adaptable. Besides, specific musical attributes can be extracted from symbolic data, which can enable more precise control.

There are some works \cite{butter, yangyin, GPT-4} attempt to generate symbolic music from text descriptions, but they also face issues with unnatural text descriptions and poor performance. Limitations exist in certain approaches \cite{butter} where the model is restricted to specific textual inputs, hindering its ability to generalize text representations to a more user-friendly format. Besides, some work \cite{butter} can only control music generation based on limited music aspects, which limits the model's ability to capture the full range of musical creativity and does not satisfy the requirements of all users. While certain approaches \cite{yangyin, GPT-4} allow for natural language inputs, their control accuracy and musical quality is limited due to the requirement for large amounts of paired text-music data. Mubert\footnote{\url{https://mubert.com/}} can produce editable MIDI compositions, but it merely retrieves and combines pre-existing music pieces rather than generating novel ideas, limiting its responsiveness to input prompts.

We propose MuseCoco, a system for generating symbolic music from text descriptions by leveraging musical attributes. 
MuseCoCo, stands for Music \textbf{Co}mposition \textbf{Co}pilot, is a powerful tool that empowers musicians to generate music directly from provided text descriptions. By harnessing the capabilities of MuseCoCo, musicians experience a substantial increase in efficiency, eliminating the need to create music entirely from scratch.
Our approach breaks down the task into two stages: text-to-attribute understanding and attribute-to-music generation stage. Musical attributes can be easily extracted from music sequences or obtained from existing attribute-labeled datasets, allowing the model in the attribute-to-music generation stage to be trained in a self-supervised manner. Musical attributes can also be inferred from natural languages in the text-to-attribute understanding stage. To synthesize paired text-to-attribute data, templates are created for each attribute, and a subset of these templates is combined and further refined into a coherent paragraph using ChatGPT's language generation capabilities.

With self-supervised training in attribute-to-music generation and supervised learning via the help of synthesized paired data in text-to-attribute understanding, a large amount of symbolic music data can be leveraged without the need of providing textual prompts manually. This can help improve model performance by increasing data amount and model size simultaneously. Besides, by leveraging various music attributes, explicit control can be achieved in generating music across multiple aspects, enabling fine-grained manipulation and customization of the generated musical output. Moreover, due to the two-stage design, MuseCoco can support multiple ways of controlling. For instance, musicians with a strong knowledge of music can directly input attribute values into the second stage to generate compositions, while users without a musical background can rely on the first-stage model to convert their intuitive textual descriptions into professional attributes. Thus, MuseCoco allows for a more inclusive and adaptable user experience than those systems that directly generate music from text descriptions.

The main contributions of this work are as follows: 
\begin{itemize}[leftmargin=*]
    \item We introduce MuseCoco, a system that seamlessly transforms textual input into musically coherent symbolic compositions. This innovative approach empowers musicians and general users from diverse backgrounds and skill levels to create music more efficiently and with better control.
    \item With this two-stage framework, a large amount of symbolic data can be used without the need for labeled text descriptions. It offers users two engagement options: directly specifying attribute values or leveraging text descriptions to control the music generation process.
    \item Subjective evaluation results have demonstrated that MuseCoco outperforms baseline systems in terms of musicality, controllability, and overall score, achieving a minimum improvement of 1.27, 1.08, and 1.32, respectively. Additionally, there is a significant boost of about 20\% in objective control accuracy, further affirming the system's enhanced performance.
    \item We also extend our model to a large scale with 1.2 billion parameters, which exhibits notable controllability and musicality, further enhancing its performance.
\end{itemize}

\section{Related Work}
\subsection{Text-to-Music Generation}

Text-to-music generation based on deep learning has been an active research area.
Though paired text-music data is scarce, there have been many works for generating audio music from input prompts.
A database containing music created by musicians and sound designers with three tags (genres, moods and activities) is constructed in MubertAI\footnote{\url{https://github.com/MubertAI/Mubert-Text-to-Music}}. It assigns the closest tags for the input prompt and generates combination of sounds from the database based on these tags.
Riffusion\footnote{\url{https://www.riffusion.com/about}} leverages stable diffusion \cite{stablediffusion} to obtain images of music spectrograms paired with input text and produce audio from them. Meanwhile, both Moûsai \cite{museai} and ERNIE-Music \cite{Erniemusic} apply diffusion models with self-collected text-audio datasets to achieve text-audio music generation.
Recently, \citet{MuLan} create a large audio-text dataset to train a joint embedding model linking music audio and natural language music descriptions, MuLan, which helps address the absence of text-music paired data.
For example, in Noise2Music \cite{huang2023noise2music}, MuLan assigns captions to unlabeled audio clips to create music-text pairs.
By directly using input text representation from MuLan during inference, MusicLM \cite{agostinelli2023musiclm} can be trained to output audio music from Mulan audio representation on existing audio data to complete the text-to-music generation task without paired text-audio data.
However, the challenge with musical audio generation from texts lies in its limited editability, which poses limitations in the composition process.
By contrast, symbolic music, notated as a sequence of musical symbols, can be simply interpreted and manipulated by humans and machines.

Only a few works focus on generating symbolic music from text descriptions.
BUTTER \cite{butter}, a music-sentence representation learning framework, proposes music representation disentanglement and cross-modal alignment to firstly generate music presented by ABC notations\footnote{\url{https://abcnotation.com/}} from text including four musical keywords (the key, meter, style, and others). Limited to the folk song datasets and a few musical factors, it cannot generate symbolic music in many varieties.
Instructed with musical input prompts, the large language model GPT-4 \cite{GPT-4} can also generate ABC notation music, however, without any nontrivial form of harmony \cite{GPT-4music}.
\citet{yangyin} also explores the text-to-music capability of pre-trained language models. Though fine-tuned with more than 200k text and ABC notation music pairs, the model cannot align the musical attribute values in loose text with the generated music well enough and can only generate music with solo tracks.

Previous work directly generates music from text descriptions, which lacks explicit control over the generation process. In this paper, we propose MuseCoco, which can generate symbolic music from text descriptions with high control accuracy and good musicality. The generated music, in its symbolic format, offers easy editability and can be explicitly controlled through attribute values derived from text descriptions.

\subsection{Controllable Music Generation}
Controllable music generation refers to the ability to exert control over specific aspects or characteristics of the generated music. By applying conditional generative models (such as conditional VAE\cite{wang2020learning, tan2020music, vonfigaro}, GAN\cite{neves2022generating, zhu2022quantized} or diffusion models\cite{huang2023noise2music, Erniemusic}), previous work leverage different conditions to generate music: Some leverage descriptions like emotion \cite{EMOPIA2021,ferreira2022controlling, bao2022generating}, style \cite{mao2018deepj, wang2022cps, choi2020encoding} or structure\cite{museformer,zhang2022structure} as conditions to control music generation. The other applies music descriptions such as instrumentation\cite{ens2020mmm, Di2021Video}, chord\cite{wang2020learning, wu2022chord}, note density \cite{tan2020music, vonfigaro}, etc. to generate music.  

However, general users usually have multiple requirements for generated music, and previous work with limited control over specific music aspects may result in limited expressiveness and diminished adaptability to different musical contexts. 
Besides, previous work can only control music generation with specified music attributes, which restricts the model ability to convey complex and nuanced musical ideas or concepts that can be effectively communicated through textual descriptions. Text-based input is easily accessible and familiar to users, making it a practical and widely adopted choice for guiding generative tasks, such as text-to-image\cite{stablediffusion, ramesh2022hierarchical, saharia2022photorealistic, ramesh2021zero} and text-to-music generation\cite{huang2023noise2music, agostinelli2023musiclm, yangyin, GPT-4music}. Previous work has struggled to generate symbolic music directly from textual descriptions provided by users. 
MuseCoco maps text input to music attributes and then utilizes music attributes to control music generation, which can effectively solve the above problems.

\section{MuseCoco}

\subsection{Overview}

\begin{figure}
    \centering
\includegraphics[width=0.95\textwidth]{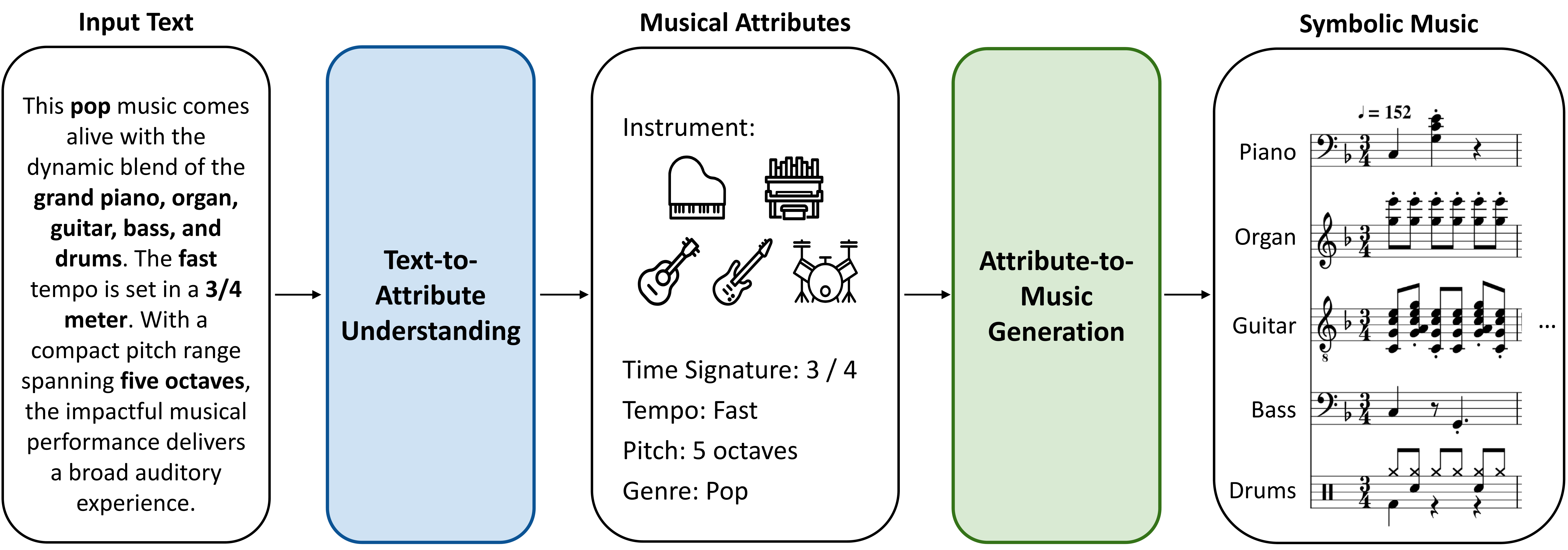}
    \caption{The two-stage framework of MuseCoco. Text-to-attribute understanding extracts diverse musical attributes, based on which symbolic music is generated through the attribute-to-music generation stage.}
    \label{fig:pipline}
\end{figure}

To achieve text-to-music controllable generation, MuseCoco incorporates natural language and symbolic music into a two-stage framework that separates text-to-attribute understanding and attribute-to-music generation, which are trained independently.
The pipeline of MuseCoco is shown in \Cref{fig:pipline}.
In this section, we will elaborate on their technical design and model architectures respectively.

\subsection{Attribute-to-Music Generation} \label{sec:attribute_to_music_generation}

Most musical attributes can be easily obtained by extracting from music sequences (\Cref{sec:experiment}), so the music generation model in the attribute-to-music stage can be trained in a self-supervised way. This method can leverage large amounts of unlabeled data, making it a highly data-efficient approach.
Musical attributes (\Cref{tab:attributes}) can be classified into objective attributes like the tempo and meter and subjective attributes like the emotion and genre \cite{ cook2001music}. Objective attributes refer to quantifiable and measurable characteristics of musical elements, so they can be extracted from music sequences with pre-defined rules (please refer to \Cref{sec:experiment} for more details). Subjective attributes refer to the qualities or characteristics of music that are based on personal interpretation, perception, or emotional response, which can be obtained from existing attribute-labeled datasets. 
After obtaining attributes from music sequences, we append those attribute tokens as prefix tokens into music sequences to provide explicit control over the music, which makes it easier to interpret and understand how the attributes are influencing the music. Different attribute tokens can be combined or sequenced to achieve complex musical expressions and transformations. 

\begin{table}[b]
  \centering
  \caption{Musical attribute descriptions.}
  \label{tab:attributes}
  \resizebox{.97\columnwidth}{!}{
    \begin{tabular}{lll}
    \toprule
    Type  & Attribute & Description \\
    \midrule
    \multirow{9}[2]{*}{Objective} & Instrument & played instruments in the music clip\\
          & Pitch & \multicolumn{1}{p{33.285em}}{the number of octaves covering all pitches in one music clip} \\
          & Rhythm Danceability & whether the piece sounds danceable \\
          & Rhythm Intensity & the intensity of the rhythm \\
          & Bar   & the total number of bars in one music clip \\
          & Time Signature & the time signature of the music clip \\
          & Key   & the tonality of the music clip \\
          & Tempo & the tempo of the music clip \\
          & Time  & the approximate time duration of the music clip \\
    \midrule
    \multirow{3}[2]{*}{Subjective} & Artist & the artist (style) of the music clip\\
          & Genre & the genre of the music clip \\
          & Emotion & the emotion of the music clip \\
    \bottomrule
    \end{tabular}}
  
\end{table}

\begin{figure}
	\vspace{-0mm}\centering
	\begin{tabular}{c c}
		\hspace{-3mm}
            \includegraphics[height=4cm,width=6.1cm]{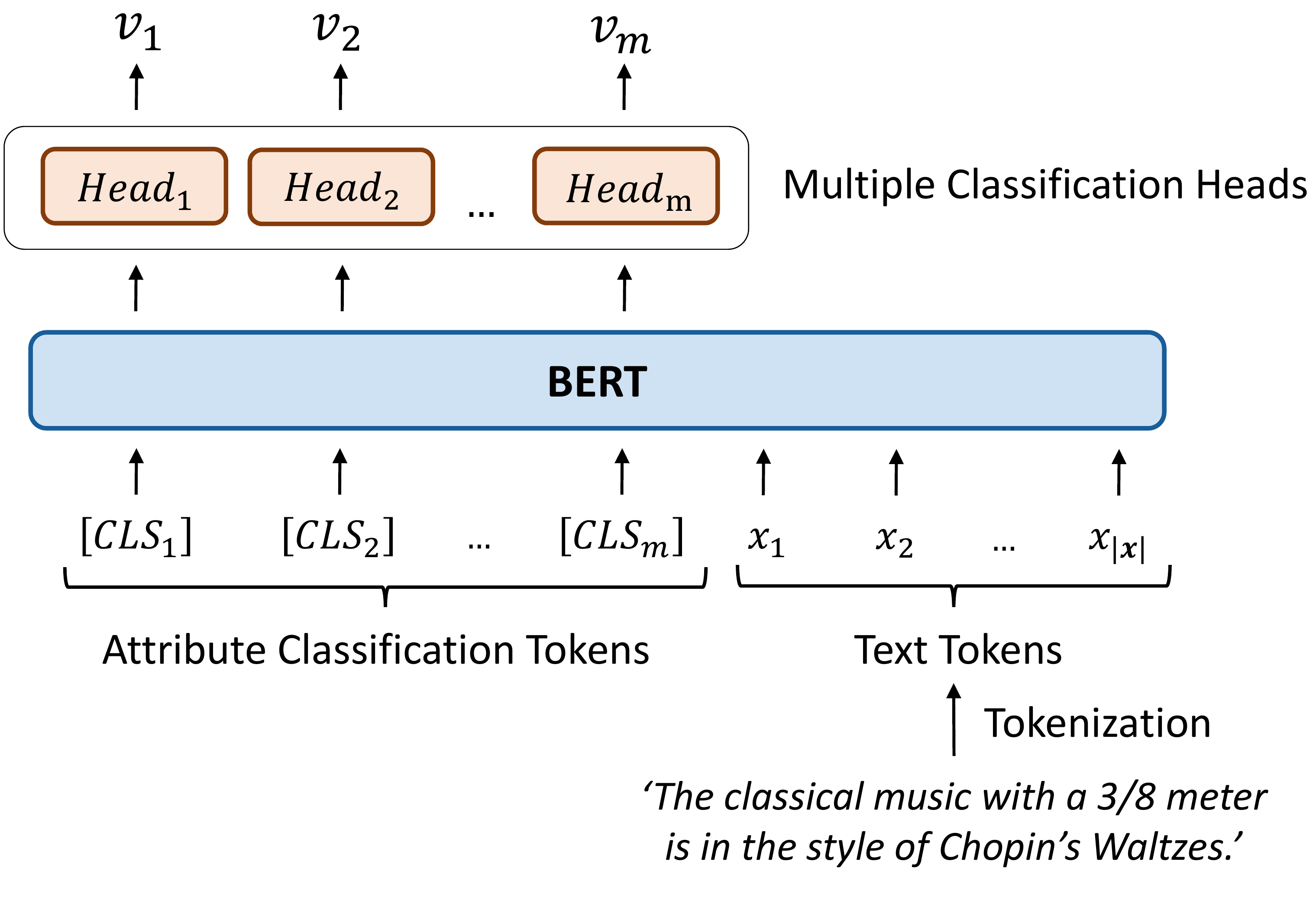} & 
		\hspace{5mm}
            \includegraphics[height=4cm,width=7cm]{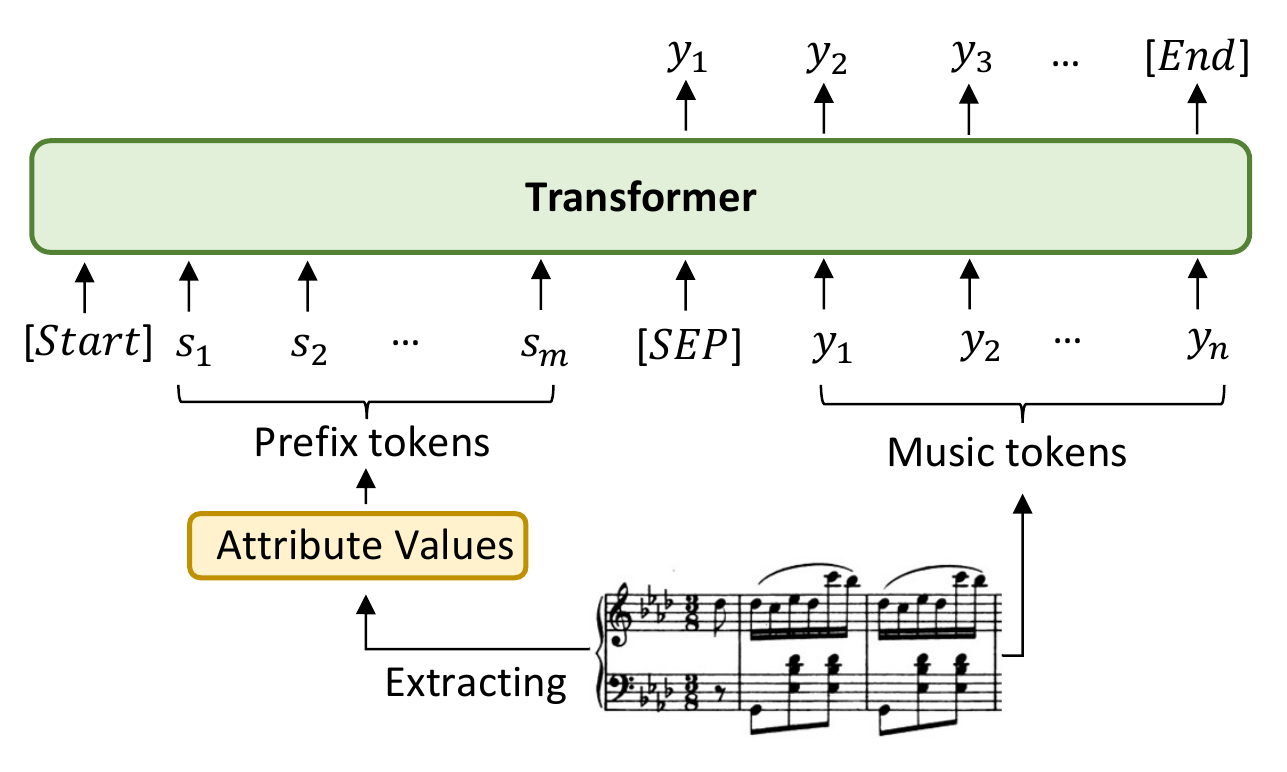} \\
		\vspace{0mm}
		\hspace{-6mm}
		(a) Text-to-attribute understanding \hspace{3mm} & 
		(b) Attribute-to-music generation   
	\end{tabular}
	\vspace{-0mm}
	\caption{During training, each stage is independently trained. During inference, the text-to-attribute understanding stage firstly extracts music attribute values, based on which the attribute-to-music generation stage secondly generates symbolic music.
	 }
	\vspace{-2mm}
	\label{fig:two_stage}
\end{figure}

Specifically, given dataset $\{ \mathcal{V}, \mathcal{Y}\}$, where  $\mathcal{Y}$ is the symbolic music dataset and $\mathcal{V}$ is the set of attribute values, MuseCoco transforms attribute values into prefix tokens to control music generation, as shown in \Cref{fig:two_stage}(b). Using prefix tokens to guide the generation process is an effective method for directing the output towards a particular direction \cite{keskar2019ctrl, li2021prefix, liu2023pre, brown2020language}. 
We pre-define $m$ musical attributes and their values as shown in \Cref{app:attribute}.
For each music sequence $\boldsymbol{y}=[y_1, y_2, ..., y_n] \in \mathcal{Y}$, and its attribute values $\boldsymbol{v} = \{v_1, v_2, ..., v_m\} \in \mathcal{V}$, we consider the following distribution:
\begin{equation}
    p(\boldsymbol{y}|\boldsymbol{v})= \prod_{i=1}^{n}p(y_i | y_{<i}, v_1, v_2,..., v_m).
\end{equation}
We encode each attribute value $v_j$ into a prefix token $s_j$. Position embeddings for such prefix tokens $\{s_j\}_{j=1}^m$ are not used. 
Since it is not usual for users to provide all attribute values in real-world scenarios, we introduce a special token, represented by $s^{NA}_j$, to exclude this attribute (i.e. $v_j$) that is not specified in inputs from influencing the music generation process. Then the input sequence is encoded into:
\begin{equation}
    s_1, s_2, s_3, ..., s_m, \texttt{[SEP]}, y_1, y_2, ..., y_n.
\end{equation}
During training, some attribute tokens (e.g., $s_2$, $s_3$) are randomly replaced with special tokens (i.e., $s^{NA}_j$) to enable adaptation to various attribute combinations. 
During inference, attribute values are provided directly from inputs. Any attributes that are absent in the inputs are represented by special tokens, which are combined with the other prefix tokens to effectively control the music generation process as required.

\subsection{Text-to-Attribute Understanding}\label{sec:text_to_attribute_understand}
Control information within input text comprises different musical attributes.
Therefore, the text-to-attribute task is required to extract musical attribute values from plain text. These attribute values will be used in the attribute-to-music generation stage to generate desired music.
As shown in \Cref{fig:two_stage}(a), this stage can be denoted as $\{\mathcal{X}, \mathcal{V}\}$, where $\mathcal{X}$ is the input text set, $\mathcal{V}$ is the value set of $m$ pre-defined musical attributes.
In the dataset, each instance $\boldsymbol{x} \in \mathcal{X}$ is paired with a combination of $m$ attribute values $\boldsymbol{v} = \{v_{i}\}^m_{i=1} $, $\boldsymbol{v} \in \mathcal{V}$.
Given a pre-trained language model $\mathcal{M}$, $\text{BERT}_{\text{large}}$ \cite{BERT}, $\boldsymbol{x}$ is converted by the tokenizer of $\mathcal{M}$ into corresponding tokens $\{x_1, x_2, \ldots, x_{|\boldsymbol{x}|}\}$.
To adapt to multiple attribute classification, we prepend $m$ attribute classification tokens $\texttt{[CLS}_{i}\texttt{]}_{i=1}^m$ to input text tokens as $m$ attribute classification heads and the encoded $\texttt{[CLS}_{i}\texttt{]}$ head is used to compute the probability distribution over the $i$ class set with a softmax classifier.
Position embeddings of $\texttt{[CLS}_{i}\texttt{]}_{i=1}^m$ are not used to be consistent with the pre-training stage.
Specifically, the input $x$ will be encoded to hidden vectors:
$$
\mathbf{h}_{\texttt{[CLS}_{1}\texttt{]}}, \mathbf{h}_{\texttt{[CLS}_{2}\texttt{]}}, \ldots, \mathbf{h}_{\texttt{[CLS}_{m}\texttt{]}}, \mathbf{h}_{x_1}, \mathbf{h}_{x_2}, \ldots, \mathbf{h}_{x_{|\boldsymbol{x}|}}
$$
The probability distribution of the value of $i$-th attribute is
$
p_{i}(v_{i} | \boldsymbol{x}) = \texttt{Softmax}(\mathbf{W_ih}_{\texttt{[CLS}_{i}\texttt{]}}+\mathbf{b_i})
$,
where $\mathbf{W_i}$ and $\mathbf{b_i}$ are learnable parameters.
The cross-entropy loss of $i$-th attribute is
$$
\mathcal{L}_{i} = -\frac{1}{|\mathcal{X}|}\sum_{\boldsymbol{x} \in \mathcal{X}} \log p_{i}(v_i | \boldsymbol{x})
$$
During training, all learnable parameters are fine-tuned by minimizing the sum of attribute cross-entropy losses ${\mathcal{L} = \sum_{i=1}^{m} \mathcal{L}_{i}}$ on $\{\mathcal{X}, \mathcal{V}\}$.

\subsection{Data Construction}
\label{sec:dataset_construction}

By using musical attributes to break down the text-to-music generation task into the attribute-to-music generation stage and the text-to-attribute understanding stage, we can leverage large amounts of symbolic music data without text descriptions. 
In the attribute-to-music generation stage, attributes can be extracted from music sequences with rules or obtained from attribute-labeled datasets (detailed in \Cref{sec:experiment}). The system only requires paired data in the text-to-attribute stage, we synthesize these text-attribute pairs in the following steps:

\begin{enumerate}[leftmargin=*]
    \item \textbf{Write templates for each attribute: }As shown in \Cref{tab:templates}, we write several templates as a set for each attribute, where its values are represented with a placeholder. By utilizing this placeholder, we can accommodate diverse combinations of attribute values without requiring exact values.
    \item \textbf{Create attribute combinations and concatenate their templates as paired texts:} 
    The generation process is usually controlled by multiple attributes together.
    Hence, constructing various different combinations of attribute values and paired text is necessary.
    Because of the long-tail distribution per attribute value in real-world data, to enrich the diversity of paired text-attribute training data and avoid the long-tailed issue, we stochastically create $\boldsymbol{v}$ per instance based on pre-defined musical attributes and their values on our own to ensure the number of instances including $v_{i}$ is balanced, i.e., each value of each attribute is sampled equally.
    And then, the paired texts of these created combinations are synthesized by simply concatenating their corresponding templates, randomly chosen from template sets of each attribute.
    \item \textbf{Refine concatenated templates via ChatGPT:}
    Since simply concatenated templates are less fluent than real users' input, ChatGPT\footnote{\url{https://chat.openai.com/}} is utilized to refine them as shown in \Cref{tab:templates}. 
    \item \textbf{Fill in placeholders: }Finally, attribute values or their synonyms fill in placeholders, ensuring that the resulting text effectively conveys the intended meaning and maintains a consistent narrative structure.
\end{enumerate}
Through these steps, we can independently construct the datasets for either of the two stages without the need for paired text-music data. 

\begin{table}
\centering
\caption{An example of synthesizing a text-attribute pair. We randomly select a template from the available templates for each attribute. Here shows two of each. Then the templates are refined by ChatGPT and then filled in values.}
\label{tab:templates}
\scalebox{0.9}{
\begin{tabular}{ccl}
\toprule
\textbf{Attribute} & \textbf{Value} & \textbf{Template} \\
\midrule
Key & Major & \multicolumn{1}{p{27em}}{This music is composed in the [KEY] key.\newline{}This music's use of [KEY] key creates a distinct atmosphere.} \\
Bar & 13 \textasciitilde{} 16 & \multicolumn{1}{p{27em}}{The song is composed of approximately [NUM\_BARS] bars.\newline{}The song comprises [NUM\_BARS] bars.} \\
Emotion & Happiness & \multicolumn{1}{p{27em}}{The music is imbued with [EMOTION]. \newline{}The music has a [EMOTION] feeling.}\\
Time Signature & 4 / 4 & \multicolumn{1}{p{27em}}{The [TIME\_SIGNATURE] time signature is used in the music.\newline{}The music is in [TIME\_SIGNATURE].} \\
\midrule 

\multicolumn{2}{p{14.5em}}{Refine via ChatGPT and fill in placeholders with values: \newline{} \{Key, Bar, Emotion, Time Signature\} }& \multicolumn{1}{p{27em}}{The music is imbued with happiness, and the major key in this music provides a powerful and memorable sound. The song progresses through 13 \textasciitilde{} 16 bars, with 4/4 as the meter of the music.}\\
\bottomrule
\end{tabular}
}
\end{table}

\section{Experiments}

\subsection{Experiment Setup}
\label{sec:experiment}

\begin{wraptable}{r}{6cm}
    \centering
    \caption{Statistics of the used datasets.}
    \vspace{0.2cm}
     \label{tab:datasets}
    \begin{tabular}{lr}
        \toprule
         Dataset &  \#MIDI \\ 
        \midrule
        MMD \cite{zeng2021musicbert} & \num{1524557} \\
        EMOPIA \cite{EMOPIA2021} & \num{1078}   \\
        MetaMidi\cite{ens2021building} & \num{612088}   \\
        POP909 \cite{wang2020pop909} & \num{909}  \\
        Symphony \cite{liu2022symphony} & \num{46360}  \\
        Emotion-gen & \num{25730} \\
        \midrule
        Total (after filtering) & \num{947659} \\
        \bottomrule
    \end{tabular}
\end{wraptable}
\paragraph{Datasets}
To train the attribute-to-music generation stage and evaluate our proposed method, we collect an assortment of MIDI datasets from online sources. \Cref{tab:datasets} lists all of the used datasets along with their respective counts of valid MIDI files. 
Specifically, the MMD dataset \cite{zeng2021musicbert} consists of many datasets
collected from the internet\footnote{We obtained the dataset for this work with the help of the authors, as it was not publicly available.}. The Emotion-gen dataset is generated by our internal emotion-controllable music generation system, and the others are all publicly released datasets. 
We did the necessary data filtering to remove duplicated and poor-quality samples, and there are \num{947659} MIDI samples remaining. From each MIDI file, we randomly extracted 3 clips within 16 bars. The attributes described in \Cref{tab:attributes} were then extracted from each clip.
The objective attribute values used in the training are directly extracted from MIDI files and the subjective attribute values are obtained from some of the datasets (details in \Cref{app:attribute}).

\paragraph{System Configuration} \label{sec:system_configuration}
For the attribute-to-music generation stage, we use a REMI-like \cite{huang2020pop} representation method to convert MIDI into token sequences. We apply Linear Transformer\cite{katharopoulos2020transformers} as the backbone model, which consists of 16 layers with causal attention and 12 attention heads. The hidden size is 1024 and FFN hidden size is 4096, yielding an approximate parameter count of 203 million. The max length of each sample is $5120$, covering at most 16-bar music segments. During training, the batch size is $64$. The dropout rate is set to $0.1$. We use Adam optimizer \cite{kingma2014adam} with $\beta_1=0.9, \beta_2=0.98$ and $\epsilon=10^{-9}$. The learning rate is  $2\times 10^{-4}$ with warm-up step $16000$ and an invert-square-root decay. In text-to-attribute understanding, we leverage $\text{BERT}_{\text{large}}$\footnote{\url{https://huggingface.co/bert-large-uncased}} as the backbone model and the max sequence length of it is set to 256, which covers common user input. We use 1,125 thousand samples for fine-tuning with a train/valid portion of 8:1. During training, the batch size is 64 and the learning rate is $1 \times 10^{-5}$.

\paragraph{Evaluation Dataset and Metrics} \label{sec:evaluation_metrics}
To evaluate MuseCoco, we construct a \textbf{standard test set} including 5,000 text-attribute pairs in the same way in \Cref{sec:dataset_construction}.
Musical attributes of each test sample originated from real music in the test set of the attribute-to-music stage, instead of creating them on our own to make sure musical rationality and all values of the attributes are covered in the test set for thorough testing.
Meanwhile, in order to accord with usual user inputs, we randomly assign the \textit{NA} value (meaning the attribute is not mentioned in the text) to some attributes per sample to synthesize text prompts with different lengths.

To evaluate the models objectively, we propose the metric called Average Sample-wise Accuracy (ASA), which is calculated by determining the proportion of correctly predicted attributes in each sample, followed by the calculation of the average prediction accuracy across the whole test set.
To conduct a subjective evaluation of MuseCoco's performance, we employ a user study.
We invite individuals with musical backgrounds to fill out the questionnaires (details in \Cref{app:user_study})
Participants are asked to rate the following metrics on a scale of 1 (lowest) to 5 (highest):
\begin{itemize}[leftmargin=*]
    \item \textbf{Musicality:} This metric assesses the degree to which the generated music exhibits qualities akin to the artistry of a human composer.
    \item \textbf{Controllability:} This metric measures how well the samples adhere to the musical attribute values specified in the text descriptions.
    \item \textbf{Overall:} This metric quantifies the overall quality of this generated music considering both its musicality and controllability.
\end{itemize}
\subsection{Comparison with Baselines}
\paragraph{Baselines}
In this study, we compare our method to two existing works for generating symbolic music from 21 text descriptions randomly selected from the standard test set:
\begin{itemize}[leftmargin=*]
\item \textbf{GPT-4}: GPT-4  \cite{GPT-4} is a large-scale language model that demonstrated its capabilities in various domains, including music.
Following \citet{GPT-4music},
we instruct GPT-4 to generate ABC notation music with the task-specific prompts (in \Cref{app:prompts}) using the official web page\footnote{\url{https://chat.openai.com/}} manually. 

\item \textbf{BART-base}: \citet{yangyin} release a language-music BART-base\footnote{\url{https://huggingface.co/sander-wood/text-to-music}}, which shows a solid performance.
Text descriptions are fed into this model and guide it to generate ABC notation music for comparison.
\end{itemize}

For a fair comparison, we convert ABC notation music generated by baselines into MIDI music using music21\footnote{\url{http://web.mit.edu/music21/}}.
As for the subjective evaluation, well-designed questionnaires including generated music from baselines and our method are distributed to individuals, who are all in music backgrounds and required to score the subjective metrics described in \Cref{sec:evaluation_metrics} (details in \Cref{app:usrmain}).
Meanwhile, to objectively compare the model ability, we calculate the average sample-wise accuracy of generated music for both baselines and our method (details in \Cref{app:objectivebaseline}).

\paragraph{Main Results}
\Cref{tab:comparison} reports the main results of the comparison.
In terms of musicality, MuseCoco achieves the highest score (mean 4.06 out of 5), indicating its ability to generate music closely resembling compositions by humans and approaching the quality of real-world music.
As for the conditional generation ability, MuseCoco outperforms all baselines in terms of controllability and the average sample-wise accuracy at 1.08 and 19.95\% respectively, which illustrates the effectiveness of controlling musical attributes with the two-stage framework.
GPT-4 also can generate more coherent music with input prompts than BART-base due to its powerful ability of language understanding.
Meanwhile, the best overall score (mean 4.13 out of 5) of MuseCoco shows our method is capable of generating the most favorite and fair-sounding music through the auditory test. Music samples generated by MuseCoco are available via this
link\footnote{\url{https://ai-muzic.github.io/musecoco/}}.

\begin{table}[htbp]
    \centering
    \caption{Comparison between MuseCoco, GPT-4, and BART-base. ASA stands for the average sample-wise accuracy.}
    \label{tab:comparison}
    \begin{tabular}{ccccc}
        \toprule
         & Musicality & Controllability & Overall & ASA (\%)\\
        \midrule
        MuseCoco & \textbf{4.06} $\pm$ 0.82 & \textbf{4.15} $\pm$ 0.78 & \textbf{4.13} $\pm$ 0.75 &  \textbf{77.59}\\
        GPT-4 \cite{GPT-4} & 2.79 $\pm$ 0.97 & 3.07 $\pm$ 1.05 & 2.81 $\pm$ 0.97 & 57.64\\
        BART-base \cite{yangyin} & 2.19 $\pm$ 1.14 & 2.02 $\pm$ 1.09 & 2.17 $\pm$ 1.03 & 31.98\\
        \bottomrule
    \end{tabular}
\end{table}

\subsection{Method Analysis}
In this section, we conduct analysis experiments on the two stages respectively.

\subsubsection{Analysis on Text-to-Attribute Understanding}
\paragraph{Attribute Comprehension}
To evaluate the ability to extract each attribute from text, we test the text-to-attribute model on the standard test set and show the classification accuracy of each attribute in \Cref{app:attribute}.
Each accuracy consistently surpasses 99\%, which proves that the model exhibits exceptional performance on all attributes and showcases the high accuracy and reliability of the text understanding ability.

\paragraph{Different classification heads} We explore the effectiveness of using multiple classification heads and report the ASA shown in \Cref{tab:head}. The model with multiple classification heads outperforms the one-head BERT by 39.87\%, which illustrates that each head can learn their corresponding attribute knowledge, and using multiple heads can improve the overall performance.

\paragraph{Different text synthetic strategies}
In order to showcase the effectiveness of the refinement strategy outlined in \Cref{sec:dataset_construction}, we evaluate its impact on enhancing the fluency and diversity of the text within the training set during the text-to-attribute stage.
We engage musicians to write 17 text descriptions manually to help evaluate and contrast the effectiveness of synthesis strategies w/ and w/o the refinement step.
As shown in \Cref{tab:refine}, we observe that fine-tuning the model with 25\% partially refined text achieves better performance than with the simply concatenated templates.

\begin{minipage}{\textwidth}
    \begin{minipage}[t]{0.55\textwidth}
    \makeatletter\def\@captype{table}
    \centering
    \caption{Analysis on multiple classification heads.}
    \label{tab:head}
    \begin{tabular}{lc}
    \toprule
          & ASA (\%) \\
    \midrule
    One Head & $60.09$ \\
    Multiple Classification Heads & 99.96 \\
    \bottomrule
    \end{tabular}
    \end{minipage}
    \begin{minipage}[t]{0.38\textwidth}
    \makeatletter\def\@captype{table}
    \centering
    \caption{Analysis on text refinement.}
    \label{tab:refine}
    \begin{tabular}{lc}
    \toprule
          & ASA (\%) \\
    \midrule
    Concatenated & 75.88 \\
    Concatenated + Refined & 78.11 \\
    \bottomrule
    \end{tabular}
    \end{minipage}
\end{minipage}

\subsubsection{Analysis on Attribute-to-Music Generation}
\label{sec:analysis_on_stage2}

\paragraph{Attribute Control Performance} 
To evaluate the controllability of the attribute-to-music generation model, we report the control accuracy results for each attribute in \Cref{app:details_on_stage2_exp}. The average attribute control accuracy is 80.42\%, demonstrating a strong capability of the model to effectively respond to the specified attribute values during the music generation process.

\paragraph{Study on Control Methods} 

We compare \textit{Prefix Control}, which is the default method of our model that uses prefix tokens to control music generation, with two other methods: 1) \textit{Embedding}: Add attribute input as embedding to token embedding; 2) \textit{Conditional LayerNorm}: Add attribute input as a condition to the layer norm layer \cite{perez2018film,chen2021adaspeech}. We utilize Musicality and average attribute control accuracy as evaluation metrics. For more details on this experiment, please refer to the description in \Cref{app:details_on_stage2_exp}. We report evaluation results in \Cref{tab:control_methods}. It is shown that \textit{Prefix Control} outperforms other methods in terms of musicality and average attribute control accuracy, with a minimum improvement of 0.04 and 19.94\% respectively, highlighting its superior capability to capture the relationship between attributes and music.

\begin{table}[htbp]
    \centering
    \caption{Comparison of different control methods. Musicality reflects the quality of the generated music. Average attribute control accuracy represents the control accuracy over all attributes, which can reflect controllability.}
    \label{tab:control_methods}
    \begin{tabular}{ccc}
        \toprule
        Method & Musicality $\uparrow$ & Average attribute control accuracy (\%) $\uparrow$ \\
        \midrule
       Embedding  & 2.97 $\pm$ 0.91 & 36.94 \\
       Conditional LayerNorm  & 3.11 $\pm$ 1.02 & 47.46 \\
       Prefix Control & \textbf{3.15} $\pm$ 1.02 & \textbf{67.40} \\
       \bottomrule
    \end{tabular}
\end{table}

\paragraph{Study on Model Size} 
We conduct a comparative analysis between two different model sizes to determine whether increasing the model size would result in improved generated results. The parameter configurations for these model sizes are presented in \Cref{tab:model_size}. The default model, referred to as \textit{large}, is the default model for the attribute-to-music generation stage. Additionally, we utilize  \textit{xlarge} model for comparison, which consists of approximately 1.2 billion parameters. The training of \textit{xlarge} model follows the same settings outlined in \Cref{sec:system_configuration}. The objective evaluation results are displayed in \Cref{tab:model_size}, which indicates that increasing the model size enhances controllability. To further evaluate the performance, we will conduct subjective listening tests to compare the subjective evaluation results for these two model sizes.
\begin{table}[htbp]
    \centering
    \caption{Comparison of different model sizes in the attribute-to-music generation stage. Average objective attribute control accuracy represents the control accuracy over objective attributes, which can reflect controllability.}
    \label{tab:model_size}
    \begin{tabular}{c|ccc|c}
        \toprule
        Model Size & Layers & $d_{model}$ & Parameters & Average objective attribute control accuracy (\%) $\uparrow$ \\
        \midrule
       large & 16 & 1024 & 203M & 83.63 \\
       xlarge  & 24 & 2048 & 1.2B & \textbf{87.15} \\
       \bottomrule
    \end{tabular}
\end{table}

\subsubsection{Comments from Musicians}
We invite professional musicians to give their comments on generated music samples from given texts based on our system. Here are some feedbacks:

\textit{The generated music closely resembles human compositions, displaying a high level of accuracy and creativity. It provides inspiration for creative compositions, showcasing interesting motifs and demonstrating skillful organization of musical elements. This work greatly improves composition efficiency, saving approximately one day of time. ---- from Musician A}

\textit{The generated music offers arrangement inspiration, exemplified by the idea of combining right-hand arpeggios and left-hand bass melody to facilitate creative expansion in composition. It sparks the concept of blending classical and popular music genres described in texts. ---- from Musician B}

\textit{The generated music significantly reduces my composition time, saving anywhere from 2 days to 2 weeks, particularly when it comes to instrumental arrangements that are outside my familiarity. The composition incorporates inspirational elements in some sections of the piece. For example, I very much enjoyed the journey through the conflicts and resolutions. In some areas towards the end it felt like I was embarking on an adventure up a mountain and through grassy fields, very interesting. ---- from Musician C}

\textit{The generated music presents the various elements of the text well, which provides great convenience for musicians to edit the music. For example, music in the style of Bach is well-created, and multiple instruments in the generated music are arranged harmoniously. Meanwhile, as for music teachers, MuseCoco is very helpful in generating desired musical examples for education. ---- from Musician and Music Teacher D}

The feedback from musicians has demonstrated our ability to enhance their workflow efficiency by reducing redundant tasks and providing creative inspiration.

\section{Conclusion}

In conclusion, this paper makes several significant contributions to the field of music generation and the application of AI in creative tasks. We introduce MuseCoco, a system that seamlessly transforms text descriptions into musically coherent symbolic compositions. This innovative approach empowers musicians and general users from diverse backgrounds and skill levels to create music more efficiently and with greater control. 

Second, we present a two-stage design that simplifies the learning process and enhances controllability. This design reduces the reliance on large amounts of paired text-music data and improves more explicit control through various aspects of attributes. By leveraging a large amount of symbolic music data, we achieve impressive musicality and establish a coherent connection between text descriptions and the generated music.

Our research demonstrates the potential of AI technologies in facilitating idea generation and streamlining the composition process for music creation tasks. By offering a powerful and adaptable tool like MuseCoco, we aim to inspire and empower artists to overcome the challenges they face in their creative pursuits and unlock new possibilities in music composition. The utilization of generative AI for creative purposes often raises concerns regarding copyright and ownership, which necessitates careful consideration moving forward. Limitations of our work can be seen in \Cref{app:limit}.

\section{Acknowledgment}
We would like to express our sincere gratitude to Amy Sorokas for her invaluable help in connecting with musicians. 
We would also like to thank musician, Grentperez, and musicians from Central Conservatory of Music in China for the collaboration and constructive feedback, which contributed to the success of this work.
Meanwhile, we sincerely appreciate all members of Wenqin Piano Society from Zhejiang University for completing the most questionnaires.

\bibliographystyle{IEEEtranN}  
\bibliography{references} 

\clearpage
\appendix
\section{Attribute Information}
\label{app:attribute}
\Cref{tab:attributes_class} shows the detailed pre-defined musical attribute values.
The value \textit{NA} of each attribute refers to that this attribute is not mentioned in the text. 
Objective attributes can be extracted from MIDI files with heuristic algorithms and subjective attributes are collected from existing datasets, as shown in Table \ref{tab:extraction}.

\begin{table}[htbp]
  \centering
  \caption{Detailed attribute values.}
   \label{tab:attributes_class}
   \resizebox{.999\columnwidth}{!}{
    \begin{tabular}{lp{40em}}
    \toprule
    Attributes & \multicolumn{1}{c}{Values} \\
    \midrule
    Instrument & 28 instruments: piano, keyboard, percussion, organ, guitar, bass, violin, viola, \newline{}cello, harp, strings, voice, trumpet, trombone, tuba, horn, brass, sax, oboe, \newline{}bassoon, clarinet, piccolo, flute, pipe, synthesizer, ethnic instrument, sound effect, drum.\newline{}Each instrument: 0: played, 1: not played, 2: NA \\
    \midrule
    \multicolumn{1}{p{6.07em}}{Pitch\newline{}Range} & \multicolumn{1}{l}{0-11: octaves,12: NA} \\
    \midrule
    \multicolumn{1}{p{6.07em}}{Rhythm\newline{}Danceability} & \multicolumn{1}{l}{0: danceable, 1: not danceable, 2: NA} \\
    \midrule
    \multicolumn{1}{p{6.07em}}{Rhythm\newline{}Intensity} & 0: serene, 1: moderate, 2: intense, 3: NA \\
    \midrule
    Bar   & \multicolumn{1}{l}{0: 1-4 bars, 1: 5-8 bars, 2: 9-12 bars, 3: 13-16 bars, 4: NA} \\
    \midrule
    \multicolumn{1}{p{6.07em}}{Time \newline{}Signature} & 0: 4/4, 1: 2/4, 2: 3/4, 3: 1/4, 4: 6/8, 5: 3/8, 6: other tempos, 7: NA \\
    \midrule
    Key   & \multicolumn{1}{l}{0: major, 1: minor, 2: NA} \\
    \midrule
    Tempo & 0: slow (<=76 BPM), 1: moderato (76-120 BPM), 2: fast (>=120 BPM), \newline{}3: NA \\
    \midrule
    Time  & \multicolumn{1}{l}{0: 0-15s, 1: 15-30s, 2: 30-45s, 3: 45-60s, 4: >60s, 5: NA} \\
    \midrule
    Artist & 0-16 artists: Beethoven, Mozart, Chopin, Schubert, Schumann,J.S.Bach, Haydn, Brahms,\newline{}Handel, Tchaikovsky, Mendelssohn,Dvorak, Liszt, Stravinsky, Mahler,Prokofiev, Shostakovich,\newline{}17: NA \\
    \midrule
    Genre & 21 genres: new age, electronic, rap, religious,international,easy listening, avant garde, \newline{}RNB, latin, children, jazz, classical, comedy, pop, reggae, stage, folk, blues, \newline{}vocal, holiday, country, symphony\newline{}Each genre: 0: with, 1: without, 2: NA \\
    \midrule
    Emotion & 0-3: the 1-4 quadrant in Russell’s valence-arousal emotion space\newline{}4: NA \\
    \bottomrule
    \end{tabular}}
\end{table}

\begin{table}[htbp]
  \centering
  \caption{Extraction methods and sources of each attributes.}
    \label{tab:extraction}
    \resizebox{.99\columnwidth}{!}{
    \begin{tabular}{lll}
    \toprule
    Type  & Attribute & Extraction Method \\
    \midrule
    \multirow{9}[2]{*}{Objective} & Instrument & directly extracted from MIDI \\
          & Pitch range & calculated based on the pitch range \\
          & Rhythm danceability & judged with the ratio of downbeat  \\
          & Rhythm intensity & judged with the average note density \\
          & Bar   & directly extracted from MIDI \\
          & Time signature & directly extracted from MIDI \\
          & Key   & judged with the note pitches based on musical rules \\
          & Tempo & directly extracted from MIDI \\
          & Time  & derived from the time signature and the number of bars \\
    \midrule
    \multirow{3}[2]{*}{Subjective} & Artist & provided by a classical music dataset in MMD \cite{zeng2021musicbert} \\
          & Genre & provided by MAGD\footnote{\url{http://www.ifs.tuwien.ac.at/mir/msd/MAGD.html}}, a classical music dataset in MMD \cite{zeng2021musicbert} and Symphony \cite{liu2022symphony} \\
          & Emotion & provided by EMOPIA \cite{EMOPIA2021} and the emotion-gen dataset \\
    \bottomrule
    \end{tabular}}
\end{table}

\section{Experiments}
\subsection{User study with baselines}
\label{app:user_study}
In the user study, participants were provided with generated music samples along with their corresponding textual prompts.
For each text description, each model (i.e., BART-base, GPT-4, MuseCoco) generated three different music clips. In each questionnaire, three samples generated with the same text conditions were randomly picked from samples generated by BART-base, GPT-4, and MuseCoco respectively as a group. Each participant was asked to evaluate 7 groups for comparison.
Three subjective metrics, musicality, controllability, and an overall score, are rated on a scale of 1 (lowest) to 5 (highest).
The participants were first requested to evaluate their music profession level, as depicted in Table \ref{tab:music_prof}. To ensure the reliability of the assessment, only individuals with at least music profession level 3 were selected, resulting in a total of 19 participants. Secondly, they were instructed to independently evaluate two separate metrics: musicality and controllability, ensuring that scoring for one metric did not influence the other.
They are also asked to give an overall score to evaluate the generated music comprehensively.
For the collected results, we computed the mean and variance for each metric. The results can be found in Table \ref{tab:comparison}.

\begin{table}[h]
\centering
\caption{Music Profession Level}
\scalebox{0.9}{
\begin{tabular}{cc}
\toprule
Level & Description \\
\midrule
1 & I rarely listen to music. \\
\midrule
2 & \multicolumn{1}{p{39.93em}}{I haven't received formal training in playing or music theory, but I often listen to music and have my preferred styles, musicians, and genres.} \\
\midrule
3 & \multicolumn{1}{p{39.93em}}{I have some basic knowledge of playing an instrument or music theory, but I haven't received formal training.} \\
\midrule
4 & \multicolumn{1}{p{39.93em}}{I haven't received formal training, but I have self-taught myself some aspects such as music theory or playing an instrument. I am at an amateur level (e.g., CCOM piano level 6 or above).} \\
\midrule
5 & I have received professional training in a systematic manner. \\
\bottomrule
\end{tabular}}
\label{tab:music_prof}
\end{table}

\subsection{Objective Comparison with baselines}
\label{app:objectivebaseline}
In this section, we introduce how to calculate the objective metric, the average
sample-wise accuracy (ASA), in \Cref{tab:comparison}.
As for MuseCoco, ten music clips are generated per prompt and we report ASA of them among the overall standard test set.
Since it is labor-intensive to leverage GPT-4 with the official web page, we only guide GPT-4 to produce five music clips per prompt and calculate the ASA of 21 prompts randomly sampled from the standard test set.
Besides, we utilize the released text-tune BART-base checkpoint\footnote{\url{https://huggingface.co/sander-wood/text-to-music}} to generate five music clips per prompt and report the ASA of 44 prompts randomly chosen from the standard test set.

\subsection{Text-to-attribute understanding}
\label{app:usrmain}
As shown in \Cref{tab:stage1main}, all attribute control accuracy is close or equal to 100\%, which indicates our model with multiple classification heads in the text-to-attribute understanding stage performs quite well.

\begin{table}[htbp]
  \centering
  \caption{Attribute control accuracy (\%) of the text-to-attribute understanding model. I: Instrument.}
  \label{tab:stage1main}
  \resizebox{.99\columnwidth}{!}{
    \begin{tabular}{lclclc}
    \toprule
    \multicolumn{1}{c}{Attribute} & Accuracy(\%) & \multicolumn{1}{c}{Attribute} & Accuracy(\%) & \multicolumn{1}{c}{Attribute} & Accuracy(\%) \\
    \midrule
    I\_piano & 100.00 & I\_clarinet & 99.92 & Genre\_comedy\_spoken & 100.00 \\
    I\_keyboard & 99.92 & I\_piccolo & 99.94 & Genre\_pop\_rock & 100.00 \\
    I\_percussion & 100.00 & I\_flute & 99.62 & Genre\_reggae & 100.00 \\
    I\_organ & 100.00 & I\_pipe & 100.00 & Genre\_stage & 100.00 \\
    I\_guitar & 99.92 & I\_synthesizer & 100.00 & Genre\_folk & 100.00 \\
    I\_bass & 99.84 & I\_ethnic\_instruments & 99.98 & Genre\_blues & 100.00 \\
    I\_violin & 99.92 & I\_sound\_effects & 99.98 & Genre\_vocal & 100.00 \\
    I\_viola & 99.96 & I\_drum & 100.00 & Genre\_holiday & 100.00 \\
    I\_cello & 99.92 & Genre\_new\_age & 99.98 & Genre\_country & 100.00 \\
    I\_harp & 100.00 & Genre\_electronic & 100.00 & Genre\_symphony & 100.00 \\
    I\_strings & 99.96 & Genre\_rap & 100.00 & Bar  & 100.00 \\
    I\_voice & 99.70 & Genre\_religious & 100.00 & Time Signature & 100.00 \\
    I\_trumpet & 99.96 & Genre\_international & 100.00 & Key    & 100.00 \\
    I\_trombone & 99.94 & Genre\_easy\_listening & 100.00 & Tempo  & 99.84 \\
    I\_tuba & 100.00 & Genre\_avant\_garde & 100.00 & Octave    & 100.00 \\
    I\_horn & 99.94 & Genre\_rnb & 100.00 & Emotion   & 99.80 \\
    I\_brass & 100.00 & Genre\_latin & 100.00 & Time   & 100.00 \\
    I\_sax & 99.84 & Genre\_children & 100.00 & Rhythm Danceability    & 100.00 \\
    I\_oboe & 99.94 & Genre\_jazz & 100.00 & Rhythm Intensity    & 99.88 \\
    I\_bassoon & 99.96 & Genre\_classical & 100.00 & Artist  & 100.00 \\
    \bottomrule
    \end{tabular}}
\end{table}

\subsection{Details of Analysis on Attribute-to-music Generation}

\label{app:details_on_stage2_exp}
\paragraph{Attribute Control Accuracy}
We report the control accuracy for each attribute on the test dataset, as shown in \Cref{tab:stage2_accuracy}. The average attribute control accuracy of 80.42\%, which provides substantial evidence for the model's proficiency in effectively controlling music generation using music attributes.
\begin{table}[htbp]
  \centering
  \caption{Accuracy (\%) of each attribute for attribute-to-music generation. I: Instrument.}
    \label{tab:stage2_accuracy}
    \begin{tabular}{lclc}
    \toprule
    \multicolumn{1}{c}{Attribute} & Accuracy(\%) & \multicolumn{1}{c}{Attribute} & Accuracy(\%) \\
    \midrule
    I\_piano & 96.20 & I\_clarinet & 90.63 \\
    I\_keyboard & 79.55 & I\_piccolo & 86.67 \\
    I\_percussion & 65.19 & I\_flute & 86.73 \\
    I\_organ & 80.55 & I\_pipe & 70.73 \\
    I\_guitar & 91.81 & I\_synthesizer & 78.28 \\
    I\_bass & 93.11 & I\_ethnic\_instruments & 77.69 \\
    I\_violin & 87.88 & I\_sound\_effects & 51.74 \\
    I\_viola & 92.03 & I\_drum & 95.96 \\
    I\_cello & 86.50 & Bar  & 71.80 \\
    I\_harp & 74.87 & Time Signature & 99.14 \\
    I\_strings & 86.08 & Key    & 57.42 \\
    I\_voice & 75.82 & Tempo  & 92.71 \\
    I\_trumpet & 84.86 & Octave    & 61.56 \\
    I\_trombone & 84.64 & Time   & 65.82 \\
    I\_tuba & 93.08 & Rhythm Danceability    & 88.04 \\
    I\_horn & 80.13 & Rhythm Intensity    & 80.47 \\
    I\_brass & 77.27 &  Genre     &  73.08 \\
    I\_sax & 81.74 &   Emotion    &  69.45 \\
    I\_oboe & 85.23 &    Artist   &  50.03 \\
    I\_bassoon & 90.72 &       &  \\
    \bottomrule
    \end{tabular}
\end{table}
\paragraph{Study on Control Methods}

To verify the effectiveness of the control method in the attribute-to-music generations stage, we compare \textit{Prefix Control} with two methods: \textit{Embedding} and \textit{Conditional LayerNorm}. For efficiency, we conducted this study on reduced-size models as follows: The backbone model of this experiment is a 6-layer Linear Transformer with causal attention. The hidden size is 512 and the FFN hidden size is 2048. The other experiment configuration is the same as \Cref{sec:system_configuration}. 
Since the control accuracy of objective attributes can be easily calculated, we only need to measure the controllability of each subjective attribute in listening tests. The control accuracy of each attribute is shown in \Cref{tab:stage2_accuracy}. Finally, the average attribute control accuracy can be calculated based on the accuracy results from both types of attributes. 
To measure the controllability of subjective attributes (such as emotion and genre), we invite 12 participants to conduct a listening test. Each participant was provided with 18 music pieces (6 pieces per control method) with corresponding subjective attributes. We asked each participant to answer: 1) Musicality(five-point scale): How similar it sounds to the music composed by a human. 2) Controllability: Does it align with the given attributes. Then we report the musicality and average attribute accuracy score in Table \Cref{tab:control_methods}. The experimental results clearly demonstrate that \textit{Prefix Control} outperforms the other two methods in terms of musicality and controllability.

\subsection{Usage of GPT models}
\label{app:prompts}

\begin{figure}
    \centering
\includegraphics[width=1.0\textwidth]{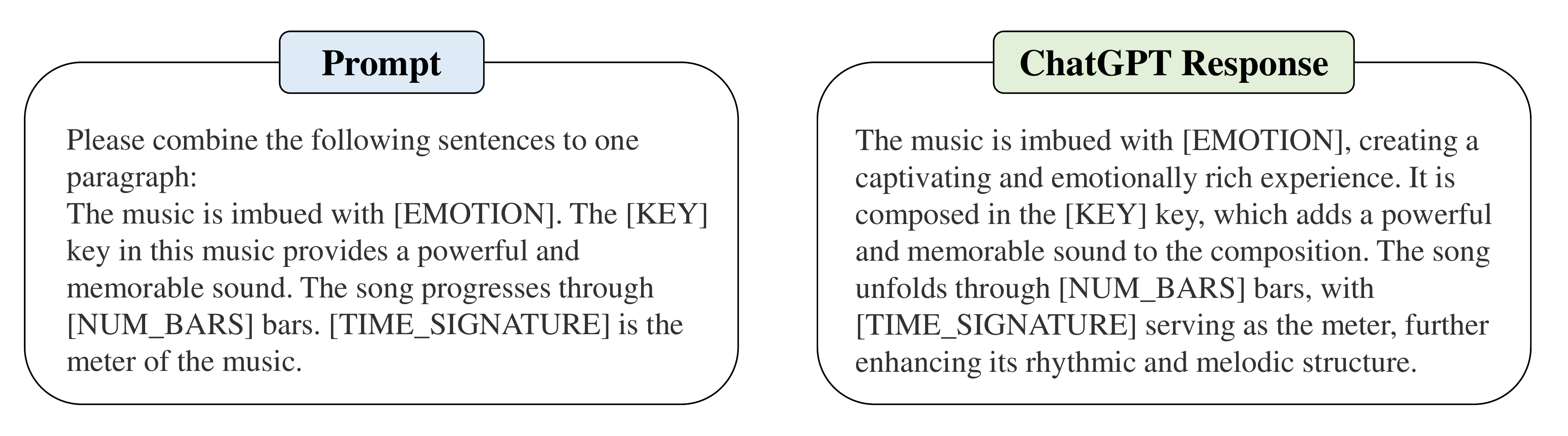}
    \caption{ChatGPT refines concatenated templates in the prompt.}
    \label{fig:ChatGPT}
\end{figure}

\paragraph{Refine texts with ChatGPT}
As shown in \Cref{fig:ChatGPT}, in order to make text descriptions more coherent and fluent, we feed concatenated templates into ChatGPT with a prompt \textit{Please combine the following sentences to one
paragraph} and then ChatGPT will give a response containing all templates within a compact paragraph.
\begin{figure}
    \centering
\includegraphics[width=1.0\textwidth]{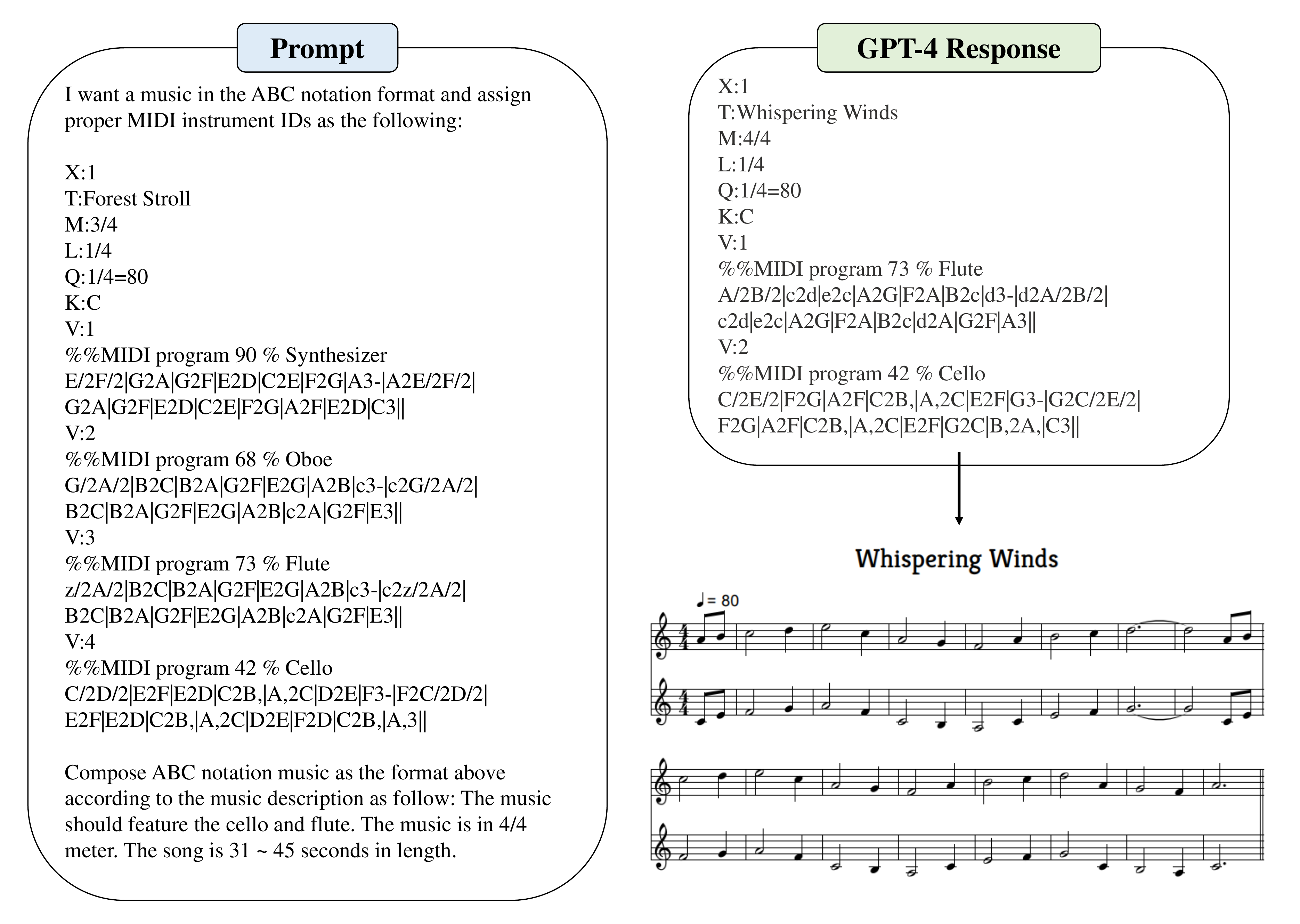}
    \caption{GPT-4 generates ABC notation tunes based on the prompt.}
    \label{fig:GPT-4}
\end{figure}

\paragraph{Generate ABC notation music with GPT-4}
To use GPT-4 as the baseline method for comparison, we design the instruction to guide GPT-4 as shown in \Cref{fig:GPT-4}.
GPT-4 can only generate symbolic music in ABC notation, so we need to explicitly point out the format. Besides, since GPT-4 can generate various ABC notation formats, some of which cannot be processed by music21, we provide an ABC notation example, teaching GPT-4 to follow its format. Meanwhile, we use the prompt, \textit{Compose ABC notation music as the format above according to the music description as follows: [text descriptions]} to let GPT-4 generate music according to the text description.
And we finally convert the ABC notations into MIDI for a fair comparison.

\section{Limitation}
\label{app:limit}

This work is mainly about generating symbolic music from text descriptions, which does not consider long sequence modeling especially. To address this, we can employ Museformer \cite{museformer} as the backbone model, which proposes fine- and coarse-grained attention for handling long sequences.

The attribute set provided in this work represents only a subset of all music attributes. We aim to further explore additional attributes to offer a broader range of control options for music generation, ensuring greater diversity in the creative process.

The possibility of regenerating music based on additional text descriptions to assist users in refining their compositions is an aspect that is worth exploring.

\end{document}